\documentclass[final,5p,times,twocolumn]{elsarticle}
\usepackage{graphicx}
\usepackage{amssymb}
\journal{Journal of Physics and Chemistry of Solids}
\begin{document}
\begin{frontmatter}
\title{Non-BCS superconductivity for underdoped cuprates by spin-vortex attraction}
\author[1]{P.A. Marchetti\corref{cor1}}\ead{marchetti@pd.infn.it}
\author[2]{F. Ye}
\author[3]{Z.B. Su}
\author[4,3]{L. Yu}
\address[1]{Dipartimento di Fisica ``G. Galilei'' and INFN,
I-35131 Padova, Italy }
\address[2]{College of Material Science and Optoelectronics
  Technology, Graduate
   University of Chinese Academy of Science,
  Beijing 100049, China}
\address[3]{Institute of Theoretical Physics, Chinese
Academy of
  Sciences,  Beijing 100190, China}
\address[4]{Institute of Physics, Chinese Academy of
Sciences,
 Beijing 100190, China}

\cortext[cor1]{}

\begin{abstract}
  Within a gauge approach to the t-J model, we propose a new, non-BCS
  mechanism of superconductivity for underdoped cuprates. The gluing
  force of the superconducting mechanism is an attraction between spin
  vortices on two different N\'eel sublattices, centered around the
  empty sites described in terms of fermionic holons. The spin
  fluctuations are described by bosonic spinons with a gap generated by
  the spin vortices. Due to the no-double occupation constraint, there
  is a gauge attraction between holon and spinon binding them into a
  physical hole. Through gauge interaction the spin vortex attraction
  induces the formation of spin-singlet (RVB) spin pairs with a lowering
  of the spinon gap.  Lowering the temperature the approach exhibits two
  crossover temperatures: at the higher crossover a finite density of
  incoherent holon pairs are formed leading to a reduction of the hole
  spectral weight, at the lower crossover also a finite density of
  incoherent spinon RVB pairs are formed, giving rise to a gas of
  incoherent preformed hole pairs, and magnetic vortices appear in the
  plasma phase. Finally, at a even lower temperature the hole pairs
  become coherent, the magnetic vortices become dilute and
  superconductivity appears.  The superconducting mechanism is not of
  BCS-type since it involves a gain in kinetic energy (for spinons)
  coming from the spin interactions.
\end{abstract}


\end{frontmatter}
\section{Introduction}
We propose a new, non-BCS mechanism of superconductivity (SC) for
hole-underdoped cuprates relying in essential way upon a
``compositeness'' \cite{masy} of the low-energy hole excitation
appearing in the spin--charge gauge approach \cite{jcmp} to the 2D {\it
  t-J} model, used to describe the CuO planes.  This ``composite''
structure involves a gapful bosonic constituent carrying spin 1/2
(spinon $z_\alpha$) and a gapless spinless fermionic constituent
carrying charge (holon $h$), supported on the empty sites. An attractive
interaction mediated by an emergent slave-particle gauge field ($A_\mu,
\mu=0,1,2$) binds them into a hole resonance. In terms of this
"composite" structure we interpret two crossovers appearing in the
normal state of cuprates that are view as ``precursors'' of
superconductivity and the recovery of full coherence of the hole at the
superconducting transition.
 
\section{"Normal" state}
To give the key ingredients of the proposed SC mechanism we start
shortly reviewing some basic features of the spin-charge gauge approach
to the normal state.  In the underdoped region of the model the
disturbance of hole doping on the antiferromagnetic (AF) background
originates spin vortices dressing the holons, with opposite chirality in
the two N\'eel sublattices (see Fig.1).

Propagating in this gas of slowly moving vortices the AF spinons,
originally gapless in the undoped Heisenberg model, acquire a finite
gap, leading to a short range AF order with inverse correlation length
\begin{equation}
\label{ms}
 m_s \approx (\delta |\log \delta|)^{1/2}.
\end{equation}
In eq.(\ref{ms}) $\delta$ is the doping concentration and the
logarithmic correction is due to the long-range tail of the spin
vortices.  Eq. (\ref{ms}) agrees with experimental data in \cite{ke}.
From the no-double occupation constraint of the $t$-$J$ model emerges
the slave-particle gauge field $A_\mu$. It is minimally coupled to holon
and spinon and it takes care of the redundant $U(1)$ degrees of freedom
coming from the spin-charge decomposition of the hole ($c_\alpha$) of
the $t$-$J$ model into spinon and holon. The dynamics of the transverse
mode of the gauge field is dominated by the contribution of the gapless
holons. Their Fermi surface produces an anomalous skin effect, with
momementum scale
 \begin{equation}
Q \approx (T k_F^2)^{1/3},
\end{equation}
the Reizer momentum, where $k_F$ is the holon Fermi momentum. As a
consequence of the $T$-dependence of the Reizer momentum, the hole
(holon-spinon) and the magnon (spinon-antispinon) resonances formed by
the gauge attraction have a strongly $T$-dependentent life-time leading
to a behaviour of these excitations less coherent than in a standard
Fermi-liquid. For the appearance of Reizer skin effect the presence of a
gap for spinons is crucial, because gapless spinons would condense at
low $T$ thus gapping the gauge field through the Anderson-Higgs
mechanism and destroying the $T$-dependent skin effect that decreases
the coherence of hole and magnon.
  
\begin{figure}[htbp]
\includegraphics[width=6cm]{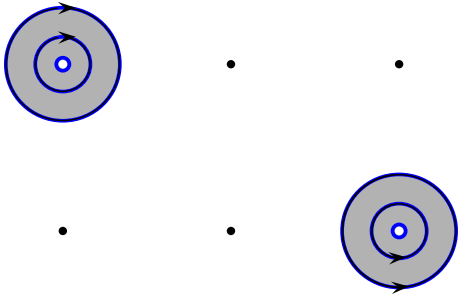}
\caption{Pictorial representation of the spin vortices dressing the
  holons represented by white circles at their center.}
\label{Fig.1}
\end{figure}
  
\section{Superconductivity mechanism}
The gluing force of the proposed superconductivity mechanism is a
long-range attraction between spin vortices centered on holons in two
different N\'eel sublattices. Therefore its origin is magnetic, but it
is not due to exchange of AF spin fluctuations as e.g. in the proposal
of \cite{pi}, \cite{sus} . Explicitely the relevant term in the
effective Hamiltonian is:
 \begin{equation}
\label{zh}
J (1-2 \delta) \langle z^* z \rangle \sum_{i,j} (-1)^{|i|+|j|} \Delta^{-1}
 (i - j) h^*_ih_i h^*_jh_j,
\end{equation}
where $\Delta$ is the 2D lattice laplacian and 
\begin{equation}
\langle z^* z \rangle \sim \int d^2q (\vec q^2
+m_s^2)^{-1} \sim (\Lambda^2+m_s^2)^{1/2}-m_s ,
\end{equation}
with $\Lambda \approx 1$ as a UV cutoff.  We propose that, lowering the
temperature, superconductivity is reached with a three-step process: at
a higher crossover a finite density of incoherent holon pairs are
formed, at a lower crossover a finite density of incoherent spinon RVB
pairs are formed, giving rise to a gas of incoherent preformed hole
pairs and a gas of magnetic vortices appears in the plasma phase, at a
even lower temperature both the holon pairs and the RVB pairs, hence
also the hole pairs, become coherent and the gas of magnetic vortices
becomes dilute. This last temperature identifies the superconducting
transition.  Clearly this mechanism relies heavily on the "composite"
structure of the hole appearing in the "normal" state.  Let us analyze
in a little more detail these three steps.
\section{Holon pairing}
 At the highest crossover temperature, denoted as
\begin{equation} 
T_{ph} \approx J (1-2 \delta) \langle z^* z \rangle ,
\end{equation}
a finite density of incoherent holon pairs appears, as consequence of
the attraction of spin vortices with opposite chirality.  We propose to
identify this temperature with the experimentally observed (upper)
pseudogap (PG) temperature, where the in-plane resistivity deviates
downward from the linear behavior.  The formation of holon pairs, in
fact, induces a reduction of the spectral weight of the hole, starting
from the antinodal region \cite{mg}. The mechanism of holon pair
formation is BCS-like in the sense of gaining potential energy from
attraction and losing kinetic energy, as shown by the reduction of the
spectral weight. As natural due to its magnetic origin, its energy scale
is however related to $J$ and not $t$, since the attraction originates
from the $J$-term of the $t$-$J$ model. We denote the BCS-like
holon-pair field by $\Delta^h$.

\section{Spinon pairing and incoherent hole pairs} 
The holon pairing alone is not enough for the appearence of
superconductivity, since its occurence needs the formation and
condensation of {\it hole} pairs.  In the previous step instead we have
only the formation of holon-pairs. One then firsty needs the formation
also of spinon-pairs.  It is the gauge attraction between holon and
spinon, that, roughly speaking, using the holon-pairs as sources of
attraction induces in turn the formation of short-range spin-singlet
(RVB) spinon pairs (see Fig.2).

\begin{figure}[htbp]
 \includegraphics[width=6cm]{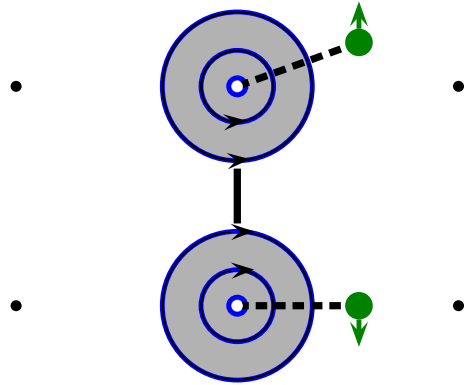}
 \caption{Pictorial representation of hole pairs, holons are represented
   by white circles surrounded by vortices, spinons by black circles
   with spin (arrow); the black line represents spin-vortex attraction,
   the dashed line the gauge attraction}
 \label{Fig.2}
 \end{figure}

 This phenomenon occurs, however, only when the density of holon-pairs
 is sufficiently high, since this attraction has to overcome the
 original AF-repulsion of spinons caused by the Heisenberg $J$-term
 which is positive in our approach, in contrast with the more standard
 RVB \cite{RVB} and slave-boson \cite{lee} approaches.  Summarizing, at
 a intermediate crossover temperature, denoted as $T_{ps}$, lower than
 $T_{ph}$ in agreement with previous remarks, a finite density of
 incoherent spinon RVB pairs are formed, which, combined with the holon
 pairs, gives rise to a gas of incoherent preformed hole pairs. We
 denote the RVB spinon-pair field by $\Delta^s$.  It turns out that for
 a finite density of spinon pairs there are two (positive energy)
 excitations, with different energies, but the same spin and
 momenta. They are given, {\it e.g.}, by creating a spinon up and
 destructing a spinon down in one of the RVB pairs. The corresponding
 dipersion relation, thus exhibits two (positive) branches (see Fig.3):
\begin{equation}
\label{sd}
\omega (\vec k) =  2t \sqrt{(m_s^2 - |\Delta^s|^2) + (|\vec k| \pm
|\Delta^s|)^2.}
\end{equation}
The lower branch exhibits a minimum with an energy lower than $m_s$,
analogous to the one appearing in a plasma of relativistic fermions
\cite{wel}; it implies a backflow of the gas of spinon-pairs dressing
the ``bare'' spinon.

\begin{figure}[htbp]
 \includegraphics[width=6cm]{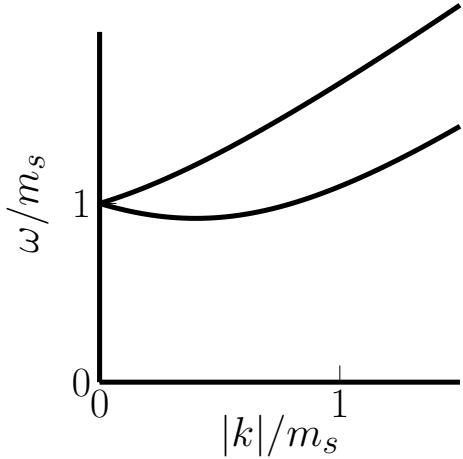}
 \caption{The positive branches of the spinon dispersion relation in
   presence of finite density of RVB spinon pairs}
\label{Fig.3} 
 \end{figure}

 Hence RVB condensation lower the spinon kinetic energy, but, as
 explained above, its occurrence needs the gauge interaction to overcome
 the spinon Heisenberg repulsion.  The two-branches dispersion of the
 spinon (\ref{sd}) is reminiscent of the hourglass shape of the neutron
 resonance found in the superconducting region and slighily above in
 temperature for underdoped samples \cite{hour}. If a suitable
 attraction mechanism for spinon and antispinon works, one can show that
 a similar dispersion is induced for the magnon resonance \cite{mg},
 directly comparable with experimental data.  As soon as we have a
 finite density of hole pairs the RVB-singlet hole-pair field $\Delta^c
 \approx \Delta^s /\Delta^h$ is non-vanishing and the gradient of its
 phase describes magnetic vortices. Hence below $T_{ps}$ a gas of
 magnetic vortice (vortex-loops in space-time) appears, in the plasma
 phase, because the incoherence of the hole pairs leads to a vanishing
 expectation value of the phase of $\Delta^c$.  Therefore, we propose to
 identify $T_{ps}$ with the experimental crossover corresponding to the
 appearance of the diamagnetic and (vortex) Nernst signal
 \cite{ong}. This interpretation is reinforced by the computation of the
 contour-plot of the spinon pair density in the $\delta-T$ plane
 \cite{mysy}, ressembling the contour-plot of the diamagnetic signal.
 The presence of holon pairs is required in advance to have RVB pairs,
 and two factors contribute to the density of holon pairs: the density
 of holons and the strength of the attraction, $\approx J (1-m_s)$ from
 (\ref{zh}). These two effects act in opposite way increasing doping,
 this yields a finite range of doping for a non-vanishing expectation
 value of $|\Delta^s|$, starting from a non-zero doping concentration,
 producing a "dome" shape of $T_{ps}$ and of the contour-plot.  The RVB
 spinon pair formation is clearly not BCS-like, the energy gain coming
 from the kinetic energy of spinons as discussed above; its energy scale
 is again related to $J$.

 \section{Hole-pair coherence and superconductivity}
 Finally, at a even lower temperature, the superconducting transition
 temperature $T_c$, both holon pairs and RVB pairs, hence also the hole
 pairs, become coherent and a d-wave hole condensate
\begin{equation}
\langle \sum_{\alpha,\beta} \epsilon_{\alpha \beta} c_{i \alpha} c_{j \beta} \rangle
\end{equation}
appears, corresponding to a non-vanishing expectation value of the
hole-pair field $\Delta^c$.  As soon as the holon and RVB pairs condense
the slave-particle global gauge symmetry is broken from $U(1)$ to ${\bf
  Z}_2$.  The Anderson-Higgs mechanism then implies a gap for the gauge
field $A_\mu$ increasing with the density of RVB pairs. In this
"Higgs"-phase the magnetic vortices become dilute.  Therefore the SC
transition appears as a 3D XY-type transition for magnetic vortices in
presence of a dynamical gauge field, similar in this respect to the
transition in the "phase-fluctuation" scenario proposed in
\cite{ek}.  One can prove that in our model a gapless gauge field is
inconsistent with the coherence of holon pairs, i.e. coherent holon
pairs cannot coexist with incoherent spinon pairs; hence the
condensation of both occurs simultaneously.  Since the gauge field
binding spinon and holon into a hole resonance becomes massive in the
superconducting state, one expects that the life-time of such resonance
becomes $T$-independent, because the $T$-dependent anomalous skin effect
appearing in the normal state is suppressed by the mass.  Therefore the
hole become coherent at the superconducting transition.

The appearance of two temperatures, one for pair formation and a lower
one for pair condensation, is typical of a BEC-BCS crossover regime for
a fermion system with attractive interaction \cite{BEC}. In this sense
the incoherent hole pairs discussed in previous section play a role
analogous to that of the "preformed pairs" considered e,g, in \cite{ue}.
In our approach, however, the gas of hole pairs appears only at finite
doping, implying a fortiory a "dome" shape for the superconductivity
region starting from a non-vanishing critical doping concentration at
$T=0$.  This result is in agreement with experimental data, but at odds
with standard fermionic BEC-BCS attractive systems, where the
condensation usually persists in the extreme BEC limit \cite{BEC}, and
with the original Mean Field version of the slave boson approach
\cite{lee}, where holon condensation occur for arbitrary small holon
density at $T=0$.
 
The non-vanishing critical doping for the "dome" exhibited in our
approach appears also in \cite{tes}, where, however, a nodal structure
is argued to be present for the hole already in the region where the
magnetic vortices are not dilute. On the contrary in that region our
approach preserves a finite FS, as consequence of the incoherence of the
holon pairs, and nodes for holes appear only below the superconducting
transition.  The superconducting transition being of XY-type is "kinetic
energy" driven, but again related to the $J$- scale since the dynamics
of vortices is triggered by the mass of the spinon and of the gauge
field, both originated from the Heisenberg term. The "kinetic energy"
driven character of the superconducting transition appears consistent
with some experimental data for underdoped and optimally doped samples
\cite{bon}.
 
\section{Final comments} 
Our approach exhibits another crossover \cite{jcmp}, 
\begin{equation}
T^* \approx t/8 \pi |\log \delta|,
\end{equation}
intersecting $T_{ps}$.  Such crossover is not directly related to
superconductivity. It corresponds to a change in the holon dispersion.
It is characterized by the emergence of a "small" holon FS around the
momenta $\pm \pi/2, \pm \pi/2$, with complete suppression of the
spectral weight for holes in the antinodal region and partial
suppression outside of the magnetic Brillouin zone. Induced physical
effects are observable experimentally both in transport and
thermodynamics \cite{ma}.  This crossover appears only in
two-dimensional bipartite lattices. Below $T^*$ the effect of
short-range AF fluctuations become stronger and the transport physics of
the corresponding normal state region ("pseudogap") is dominated by the
interplay between the short-range AF of spinons and the thermal
diffusion induced by the gauge fluctuations triggered by the Reizer
momentum, producing in turn the metal-insulator crossover \cite{jcmp}.
We identify $T^*$ in experimental data with the inflection point of
in-plane resistivity and the broad peak in the specific heat coefficient
\cite{ma}.
 
The above presentation is just a sketch of the approach, many details
have been already worked out explicitely, others remain admittedly
conjectural , but the mechanism of SC proposed here is rather complete
in its main structure and qualitatively consistent with many
experimental features, as partially discussed above.  A more complete
presentation of our approach to superconductivity will appear in
\cite{mysy}.


\begin{thebibliography}{00}
\bibitem{masy} P. A. Marchetti, A. Ambrosetti, Z. B. Su and L. Yu, J. of Phys. and Chem. of Solids {\bf 69}. 3277 (2008).
\bibitem{jcmp} P. A. Marchetti, Z. B. Su and L. Yu, J. Phys. Condens. Matter
{\bf 19}, 125209 (2007) and references therein.
\bibitem{ke}B. Keimer {\it et al.},  Phys. Rev. B {\bf 46},
14034 (1992).
\bibitem{pi} D. Pines, in Proc. Conf. on Gap Symmetry in High-Tc Superconductors,
Plenum, New York, 1998; D.J. Scalapino, J. Low Temp. Phys. {\bf 117} 179 (1999).
\bibitem{sus} M. Y. Kuchiev and O.P. Sushkov, Physica C {\bf 218}, 197 (1993);
V.V. Flambaum, M. Y. Kuchiev and O.P. Sushkov, Physica C {\bf 227},
267 (1994).
\bibitem{mg} P. A. Marchetti and M. Gambaccini, in preparation.
\bibitem{RVB} P.W. Anderson {\it et al.}, J Phys. Condens. Matter {\bf 16}, R755-R769 (2004).
\bibitem{lee} P.A. Lee  and N. Nagaosa, Phys. Rev. B {\bf 46} 5621 (1992).
\bibitem{wel} H. A. Weldon, Phys. Rev. D {\bf 40}, 2410 (1989).
\bibitem{hour} C. Stock {\it et al.} Phys. Rev. B {\bf 71}, 024522 (2005)
\bibitem{ong} Y. Wang {\it et al.}, Phys. Rev. Lett. {\bf 88}, 257003
(2002); Y. Wang, L. Li and N. P. Ong, Phys. Rev. B {\bf 73}, 024510
(2006).
\bibitem{mysy}P. A. Marchetti, F. Ye, Z. B. Su and L. Yu, to be published.
\bibitem{ek} V.J. Emery and S.A. Kivelson, Nature {\bf 374}, 434 (1995).
\bibitem{BEC} P. Nozieres, S. Schmitt-Rink, J. Low Temp. Phys. {\bf 59},
195 (1985).
\bibitem{ue} Y.J. Uemura, Physica {\bf }282, 194 (1997).
\bibitem {tes} Z. Tesanovic, Nat. Phys. {\bf 4}, 408 (2008).
\bibitem{bon}G. Deutscher {\it et al.}, Phys. Rev. B {\bf 72}, 092504
(2005);E. van Heumen {\it et al.},Phys. Rev. B {\bf 75}, 054522 (2007).
\bibitem{ma} P. A. Marchetti and A. Ambrosetti, Phys. Rev. B. {\bf 78}, 085119 (2008).
\end{thebibliography}
\end{document}